\documentclass[16pt]{article}
\usepackage{graphicx}
\usepackage{xcolor}
\usepackage[left=2.6cm, right=2.6cm, top=2.6cm, bottom=2.6cm]{geometry}
\usepackage{fancyhdr}
\usepackage[square,sort,comma,numbers]{natbib}
\usepackage{url}
\usepackage{balance}  
\usepackage{subfigure}
\providecommand{\keywords}[1]
{
  \small	
  \textbf{\textit{Keywords---}} #1
}

\title{An Empirical Study of Community Detection Algorithms on Social and Road Networks}

\author{Waqas Nawaz\\ College of Computer and Information Systems, \\Islamic University of Madinah, KSA.\\ wnawaz@iu.edu.sa}
\date{}

\begin{document}
\linespread{1.25}
\maketitle
\thispagestyle{fancy}
\setcounter{page}{185}
\section*{Abstract}
Community detection in social networks is a problem with considerable interest, since, discovering communities reveals hidden information about networks. There exist many algorithms to detect inherent community structures and recently few of them are investigated on social networks. However, it is non-trivial to decide the best approach in the presence of diverse nature of graphs, in terms of density and sparsity, and inadequate analysis of the results. Therefore, in this study, we analyze and compare various algorithms to detect communities in two networks, namely social and road networks, with varying structural properties. The algorithms under consideration are evaluated with unique metrics for internal and external connectivity of communities that includes internal density, average degree, cut ratio, conductance, normalized cut, and average Jaccard Index. The evaluation results revealed key insights about selected algorithms and underlying community structures. 

\keywords{Graph Mining, Community Detection, Road Network, Social Network, Algorithms, Community Analysis}

\section{Introduction}
Cluster, or community structure, is a grouping of graph vertices together such that intra-group edge density is higher than inter-group edge density \cite{clauset:CNM}. There are plethora of techniques for detecting communities in a network \cite{chen:MBDSGE,clauset:CNM,huang:gcluskeleton,khorasgani:topleaders,kumpula:SCP,leung:HANP,raghavan:LPA,zhao:MKMF} and it is non-trivial to decide the best approach for a certain scenario.

There are few initiatives taken in the literature to simplify this task. One of them investigated graph communities with ground-truth \cite{yang:defining}, which is not favorable in real life networks. Another made an evaluation on overlapping communities in \cite{xie2013overlapping} that is not the focus of this study. \textcolor{black}{Therefore, it} is essential to evaluate the community detection algorithms considering the inherent structural properties of graph data and choose respective metric. For instance, the study of LPA \cite{harenberg:lpaevaluation} shows poor quality results when applied on a dataset with high clustering coefficient, although as per author claim it is the best algorithm for detecting communities. Moreover, other studies \cite{yang:defining} and \cite{leskovec:empirical} say that it is necessary to use datasets with varying characteristics and corresponding metric for effective evaluation. For example, when the network contains well-separated non-overlapping communities, \emph{conductance} is the best scoring function in such cases.

The existing studies lack behind in diversity of datasets and effective evaluation measures, which we overcome in this work. We study the problem of community detection in diverse networks based on social interactions (email) and physical infrastructure(road), which have varying structural properties. It provides a validation tool to verify the correctness of the claims in literature regarding social network communities. We use unique set of measures to evaluate the resultant communities that includes internal density,
2
 average degree, conductance, cut ratio, and normalized cut. 

The key findings of this study \textcolor{black}{are} summarized as follows.
\begin{itemize}
    \item The community detection algorithms studied in this paper have shown consistent performance over social graph data in reference to existing benchmark \cite{wang:becnhmark}.
    \item We noticed a significant change in proportion of structural properties, neighborhood connectivity and degree distribution, of two social graphs, i.e. email and collaboration network.
    \item We observed few unexpected behaviors of the selected algorithms in certain cases. For example, LPA approach has shown poor performance for effectiveness (average degree) of communities on collaboration network.
\end{itemize}
 
\section{Existing Work}
In this section, we briefly elaborate the existing empirical studies on community detection algorithms and also highlight the differences to our work.

\textcolor{black}{In one of the most relevant study\cite{wang:becnhmark}, the authors have} \textcolor{black}{presented a universal framework for comparison that gives equal conditions to evaluate various community detection approaches. The authors bench-marked a set of algorithms on social networks to better analyze, evaluate, diagnose and further improve them. However, this benchmark is limited when it comes to diversity of networks and evaluation measures.} \textcolor{black}{Similarly, G. Misra's work \cite{misra2016non} analyzed eight different community detection algorithms for access control decisions in a personalized social network. This study lacks the diversity of the network under-consideration, i.e. social network, and limits the analysis of community detection algorithms to a specific application scenario.} 

\textcolor{black}{N. Grag et. al. \cite{garg2017comparative} and Z. Zhao et. al. \cite{zhao2018comparative} have studied community detection algorithms only in the context of modularity, which is one kind of measure to analyze the quality of resultant communities, with varying size of the graphs rather than the properties . Moreover, they have not considered variety of community detection algorithms. The most recent empirical study on community detection algorithms in \cite{EmpStdCommunityDetect} highlights only their advantages and disadvantages without actual analysis of community structures of the real networks. This study also overlooked the structural properties of the underlying network for resultant communities.}

\section{\textcolor{black}{Community Detection Algorithms: An Overview}}
\textcolor{black}{In this section, We briefly outline the community detection approaches under consideration for this study.}

\textcolor{black}{Community detection is a well known problem to identify and group strongly connected nodes together in a graph.} There are various approaches to detect communities in graphs and often categorized based on their nature of processing such as local (forming communities from local structures to the whole graph), global (separating out communities from the entire graph) and tree-structure (maintaining a tree with branches that represent communities) methods \cite{wang:becnhmark}. We \textcolor{black}{chose} six methods\footnote{The availability of implementation resources, provided by authors, also affected this selection process.} such that at least one method from each category for representative analysis.

\begin{itemize}
    \item \textbf{Clauset Newman Moore (CNM)} \cite{clauset:CNM} \textcolor{black}{belongs to a} hierarchical clustering \textcolor{black}{strategy} that maintains hierarchy of \textcolor{black}{the resultant} clusters. \textcolor{black}{The hierarchical clustering method follows either agglomerative or divisive approach}. \textcolor{black}{CNM uses classical agglomeration approach that is a \emph{bottom up} strategy while divisive method follows \emph{top down} strategy}. CNM optimizes the modularity of the final partition by making greedy choices. \textcolor{black}{This algorithm is effectively used for complex networks in research community, and was designed specifically to analyze the community structures of extremely large networks, i.e. millions of nodes.}

    \item \textbf{Radicchi} \cite{radicchi:defining} is also a type of hierarchical clustering algorithm but unlike CNM, it takes a divisive approach by starting from the whole graph and splitting it into communities gradually. 

    \item \textbf{Label Propagation Algorithm (LPA)} \cite{raghavan:LPA} is an efficient, near linear time, algorithm to detect community structures in large-scale networks. It is a semi-supervised algorithm that uses unlabeled nodes to find out the labels. It has an advantage in running time and performs well when there is prior information or annotated data. 

    \item \textbf{TopLeaders}\cite{khorasgani:topleaders}\textcolor{black}{, i.e. Leadership expansion algorithm,} extracts clusters from the graph identifying it as sets, consisting of a leader node and its follower nodes that are close to the leader. This algorithm requires to select initial {\it k} leaders as the number of desired communities.

    \item \textbf{Sequential Clique Percolation (SCP)} \cite{kumpula:SCP} algorithm is based on the clique percolation method and detects {\it k}-clique subgrpahs for a given value of {\it k} from dense graph by sequentially inserting edges and keeping track of the emerging community structure. \textcolor{black}{In comparison to CFinder\cite{adamcsek2006cfinder}, it finds all the cliques of single size and output the communities for all possible thresholds, while CFinder finds maximal cliques in a graph and produces communities of all possible clique sizes. Therefore, we can consider this algorithm as a alternative to CFinder. Another good thing about SCP algorithm is that it works well with large sparse graphs, however, it may not be a good option when the graph is very dense or it contains large size cliques.} 

    \item \textbf{Matrix Blocking Dense Subgraph Extraction (MB-DSGE)} \cite{chen:MBDSGE} \textcolor{black}{algorithm reorders a relatively sparse graph and extracts dense subgraphs as communities. More precisely,} for clustering, it constructs a hierarchy tree using matrix blocking technique, which groups similar columns of an adjacency matrix according to the cosine similarity measure.
\end{itemize}
 
\section{Experiments}
In this section, we explain the details of experiments including environmental setup, algorithms for comparisons, data sets, and result evaluation criteria. The detailed discussion is provided at the end.

\subsection{Environment Setup}
We analyze six representative algorithms to detect communities. These algorithms are implemented on different platforms. Our evaluation criteria is independent of any platform and considers the output results explicitly. Therefore, it is not essential to execute or implement all the algorithms on a single platform. The implementation detail for each algorithm is as follows. The default setting for each algorithm is considered for all subsequent experiments unless stated explicitly.

\begin{itemize}
	\item \textbf{CNM} (Clauset Newman Moore) is the agglomeration approach \cite{clauset:CNM}. The authors provide an executable makefile of their implementation written in C \cite{urlcode:urlcnm}. \textcolor{black}{The original version of the provided code works on unweighted and undirected graphs. However, later the authors introduce another version of the code that works on weighted and undirected graphs.}
	
	\item \textbf{Radicchi} is a divisive hierarchical clustering algorithm\cite{radicchi:defining}. The authors provide binaries written in C++ \cite{urlcode:urlradicchi} \textcolor{black}{along with the source code files.}
	
	\item \textbf{LPA} (Label Propagation Algorithm)  \cite{raghavan:LPA}. The implementation is available in R \textcolor{black}{programming} language for LPA \cite{urlcode:urllpar}. However, we also found LPA's implementation in Python language \cite{urlcode:urllpapython} and used for experiments in this study.
	
	\item \textbf{TopLeaders} algorithm gradually associates nodes to the nearest leaders and locally reelects new leaders during each iteration \cite{khorasgani:topleaders}. The authors provide an executable jar file, written in java \cite{urlcode:urltopleaders}.
	
	\item \textbf{SCP} \cite{kumpula:SCP}. \textcolor{black}{This algorithm's} source code is written in C++ and Python \textcolor{black}{\cite{urlcode:urlscp}, however,} we used Python implementation for our analysis.
	
	\item \textbf{MB-DSGE} \cite{chen:MBDSGE}. \textcolor{black}{The source code of this algorithm is implemented in C++ with the makefile which is available at \cite{urlcode:urlmbdsge}. In this implementation, the author used an open source C++ linear algebra library called Eigen \cite{urlcode:eigen}.}
\end{itemize}

\begin{figure}
	\centering 
	\includegraphics[width=0.8\textwidth]{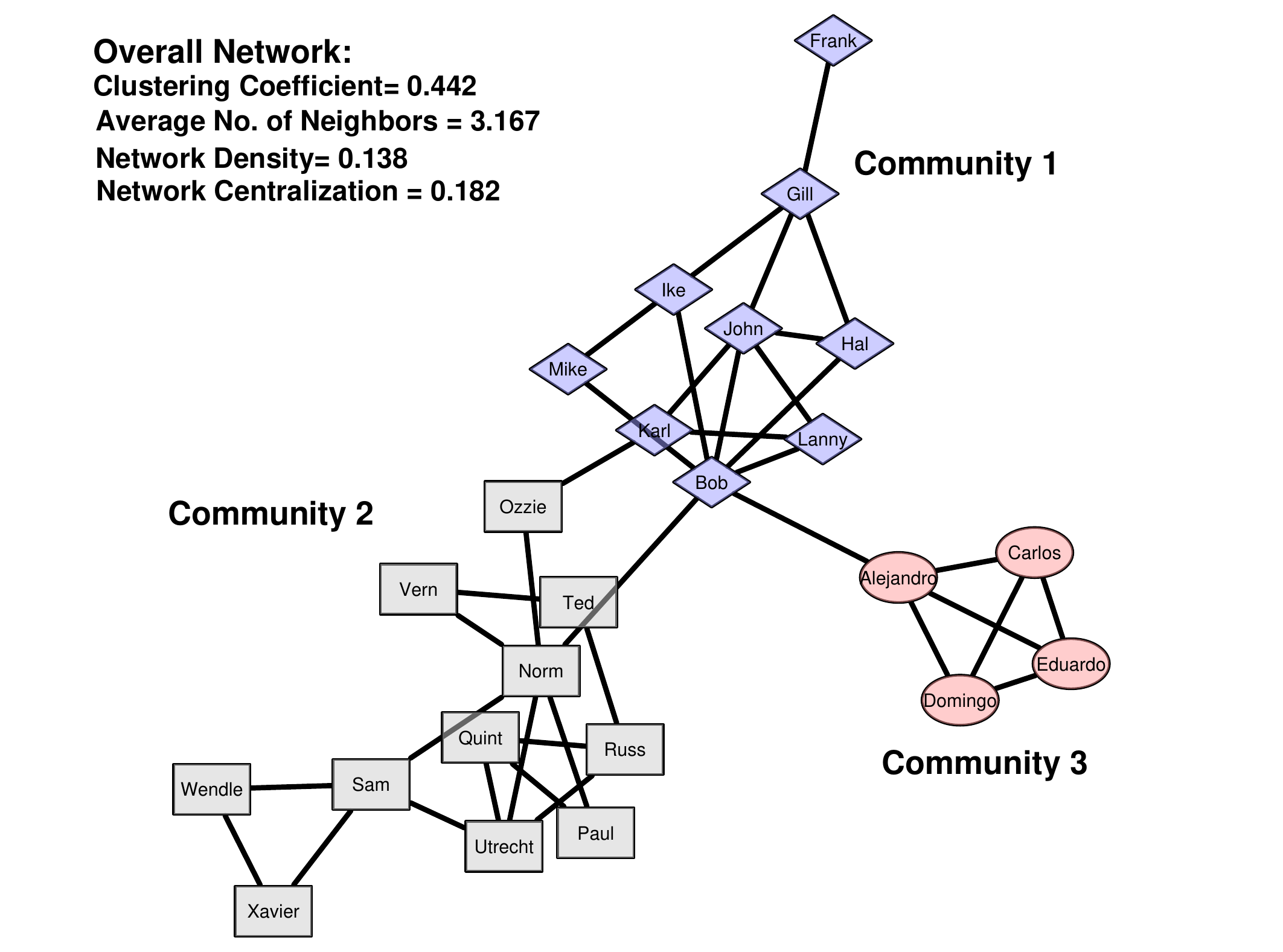}
	\caption{Strike network with communities as a ground truth.}
	\label{fig:strike}
\end{figure}

\subsection{Datasets} 
We use four widely used real-world network datasets, i.e. communication, collaboration, and road network.
\begin{itemize}
	\item \textbf{Strike}  (the communication network of employees in a sawmill): It has 24 vertices (employees), 38 edges (discussed the strike in some minimum frequency), no arcs, no loops, no line values \cite{khorasgani:topleaders}. 
	\item \textbf{HEP-PH} (High Energy Physics - Phenomenology):  This collaboration network is from the e-print arXiv and covers scientific collaborations between authors papers submitted to High Energy Physics - Phenomenology category. If an author i co-authored a paper with another author j, the graph contains a undirected edge from vertex i to vertex j. If the paper is co-authored by k authors, this generates a completely connected (sub)graph of k vertices. It contains 12008 vertices and 118521 edges \cite{snapnets}.
	\item \textbf{Enron} (email network) Enron email communication network \cite{leskovec2008statistical} covers all the email communication within a dataset of around half million emails. This data was originally made public, and posted to the web, by the Federal Energy Regulatory Commission during its investigation. Nodes of the network are email addresses and if an address i sent at least one email to address j, the graph contains an undirected edge from i to j.
	\item \textbf{roadNet-PA} (road network) This is a road network of Pennsylvania \cite{leskovec2009community}. Intersections and endpoints are represented by nodes, and the roads connecting these intersections or endpoints are represented by undirected edges.
\end{itemize}

\begin{figure}
	\centering 
	\includegraphics[width=0.8\textwidth]{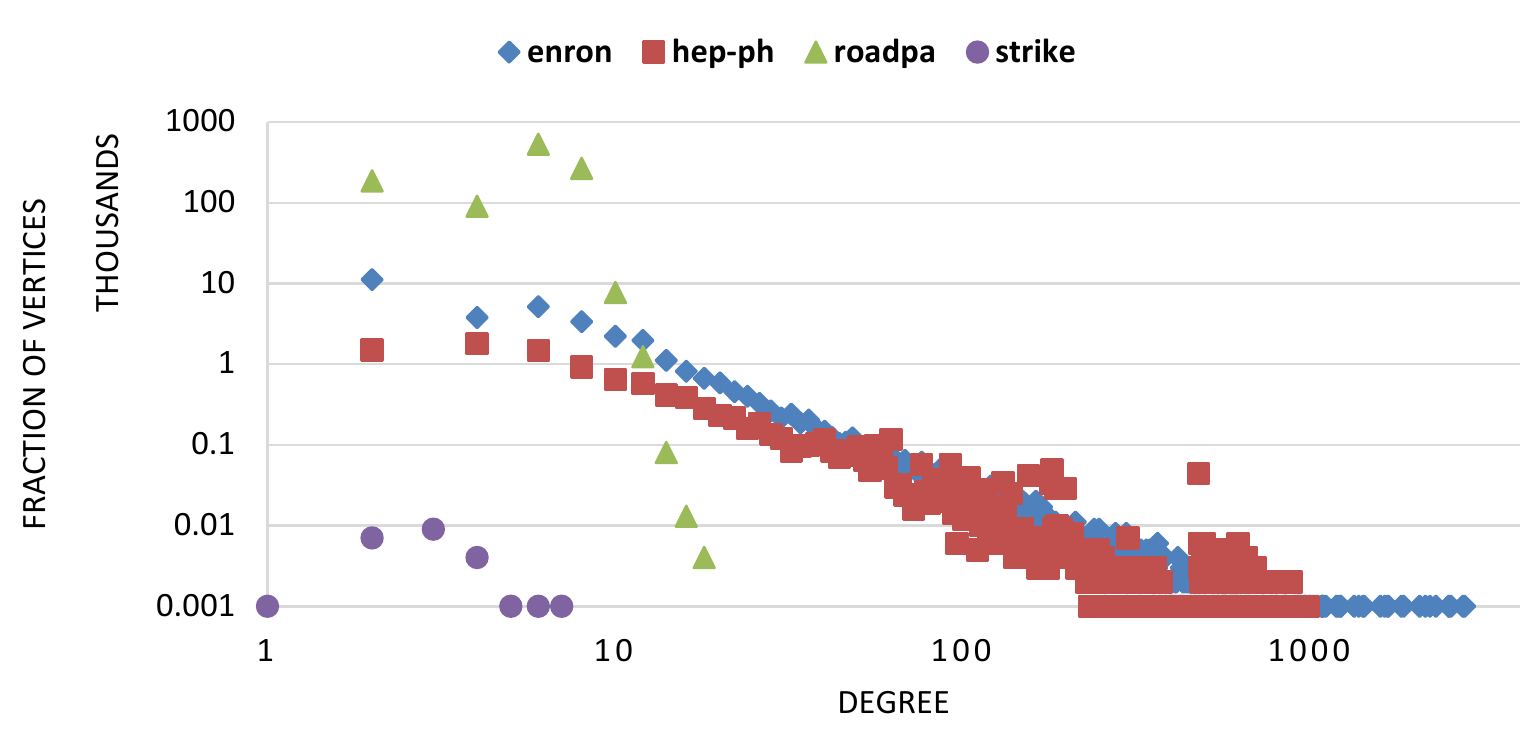}
	\caption{Degree distribution of collaboration, communication, and road networks.}
	\label{fig:degree_dist}
\end{figure}

We analyze the aforementioned datasets to highlight the diversity of these networks. It helps us to anticipate the nature of results with given knowledge about original networks. The Strike network is visualized in Figure \ref{fig:strike} with communities as a ground truth. We also computed the overall network properties for understanding, e.g. cluster coefficient is 0.44 that shows tendency towards better communities. We have shown the degree distribution of all datasets in Figure \ref{fig:degree_dist} to analyze their structural aspects. The road network has limited variations for vertex degree and distribution is somehow different from communication and collaboration networks.

\begin{figure}%
	\centering%

	\subfigure[][]{%
	\centering
	\label{fig:graph-nConnecta}%
	\includegraphics[width=0.5\textwidth]{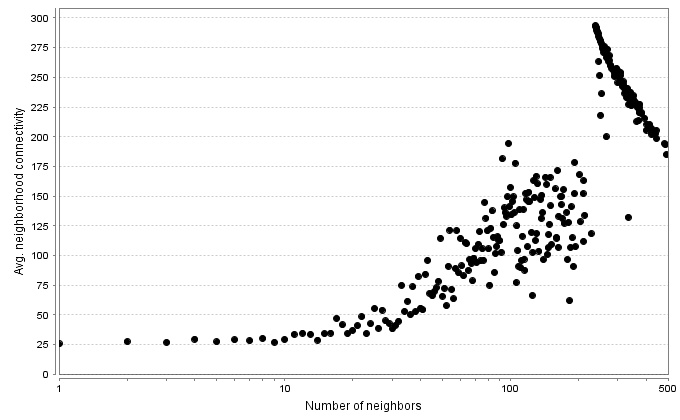}}%
	\subfigure[][]{%
	\centering
	\label{fig:graph-nConnectb}%
	\includegraphics[width=0.5\textwidth]{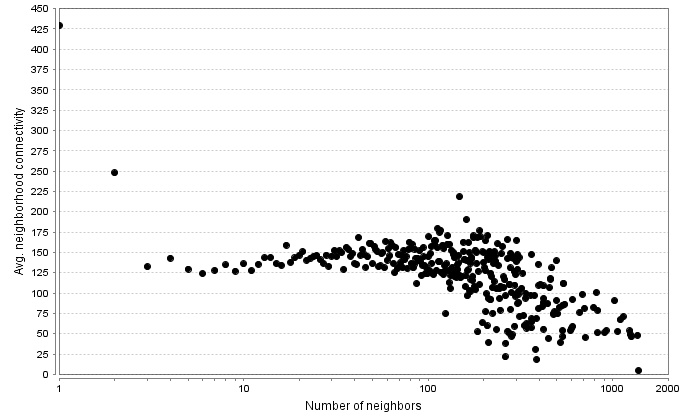}}%
	\caption{Graph neighborhood connectivity for (a) collaboration and (b) email networks. The number of neighbors are plotted on x-axis and y-axis shows the average connectivity among neighbors.}%
	\label{fig:graph-nConnect}%
\end{figure}

The neighborhood connectivity is critical for graph processing and have more tendency to discover communities. Social networks usually have high neighborhood connectivity that may produce good quality clusters. On the other hand, relatively sparse graphs have lower connectivity among neighbors of vertices, which may produce poor quality clusters due to low edge density. We plotted the neighborhood connectivity with respect to number of neighbors for collaboration (hep-ph) and communication (Enron) networks in Figure \ref{fig:graph-nConnect}. We observed a considerable difference for neighborhood connectivity between these two networks, e.g. when we consider more than 100 neighbors then the connectivity is higher for collaboration network compared to email network. It is strongly related to the size of resultant communities.  

\subsection{Evaluation criteria}
We evaluate algorithms in terms of effectiveness, accuracy and outliers. 
For effectiveness, we use scoring functions, defined in Section \ref{sec:scoreFunc}, based on internal connectivity (internal density, average degree), external connectivity (cut ratio), and metrics that combine internal and external connectivity (conductance, normalized cut). The intuition behind conductance is that a community is a set of nodes strongly connected internally than externally and other metrics following similar intuition are also popular in \textcolor{black}{research} community \cite{leskovec:empirical}. For accuracy we chose Jaccard Index that is a widely used similarity measure. It is more sensitive to overcome the small variance of the cross common fraction\textcolor{black}{, i.e.} when nodes from several different communities in one result join together as a single community in another result \cite{raghavan:LPA}.

\subsection{\textcolor{black}{Community Scoring Functions}}\label{sec:scoreFunc}
\textcolor{black}{The basic intuition for all scoring functions is that communities are sets of nodes with many connections between the members and few connections from the members to the rest of the network. Given a set of nodes S, it is considered a function $f(S)$ that characterizes how community-like is the connectivity of nodes in $S$. Let G(V,E) be an un-directed graph with $n = |V|$ nodes and $m=|E|$ edges. Let S be the set of nodes, where $n_s$ is the number of nodes in S, $n_s = |S|$; $m_s$ the number of edges in S, $m_s = |{(u, v) \in E :u \in S, v \in S}|;$ and $c_s$, the number of edges on the boundary of S, $c_s=|{(u,v) \in E : u \in S, v \notin S}|$; and $d(u)$ is the degree of node $u$.} 

\begin{itemize}
	\item \textcolor{black}{\textbf{Conductance}, $f(S)=\frac{c_s}{2m_s+c_s}$. It measures the fraction of total edge volume that points outside the cluster. A good community should have high cohesiveness (high internal conductance) as it should require deleting many edges before the community would be internally split into disconnected components. Conductance captures a notion of "surface area-to-volume" and thus it is widely-used to capture quantitatively the gestalt notion of a good network community as a set of nodes that has better internal- than external-connectivity \cite{shi2000normalized}.}
	
	\item \textcolor{black}{\textbf{Internal density}, $f(S)=\frac{m_s}{n_s(n_{s}-1)/2}$, is the internal edge density of the node set S \cite{radicchi:defining}.}
	
	\item \textcolor{black}{\textbf{Average degree}, $f(S)=\frac{2m_s}{n_s}$, is the average internal degree of the members of S \cite{radicchi:defining}.}
	
	\item \textcolor{black}{\textbf{Cut ratio}, $f(S)=\frac{c_s}{n_s(n-n_s)}$, is the fraction of existing edges (out of all possible edges) leaving the cluster \cite{fortunato2010community}.}
	
	\item \textcolor{black}{\textbf{Normalized cut}, $f(S)=\frac{c_s}{2m_{s}+c_s}+\frac{c_s}{2(m-m_{s})+c_s}$ \cite{shi2000normalized}.}	
	
	\item \textcolor{black}{\textbf{Jaccard Index} is a widely used similarity measure. It can be defined as $\frac{P_s}{P_{s}+P_{s1}+P_{s2}}$ where $P_s$ stands for the number of node pairs that are respectively classified into the same community in both results, $P_{s1}$ stands for the number of node pairs appearing in the same community in the algorithm-produced results, but in different communities based on the ground truth, and $P_{s2}$ vice versa \cite{fortunato2010community}.}
\end{itemize}

\begin{figure}%
	\centering%
	\subfigure[][]{%
	\centering
	\label{fig:strike_hepph_statsa}%
	\includegraphics[width=0.50\textwidth]{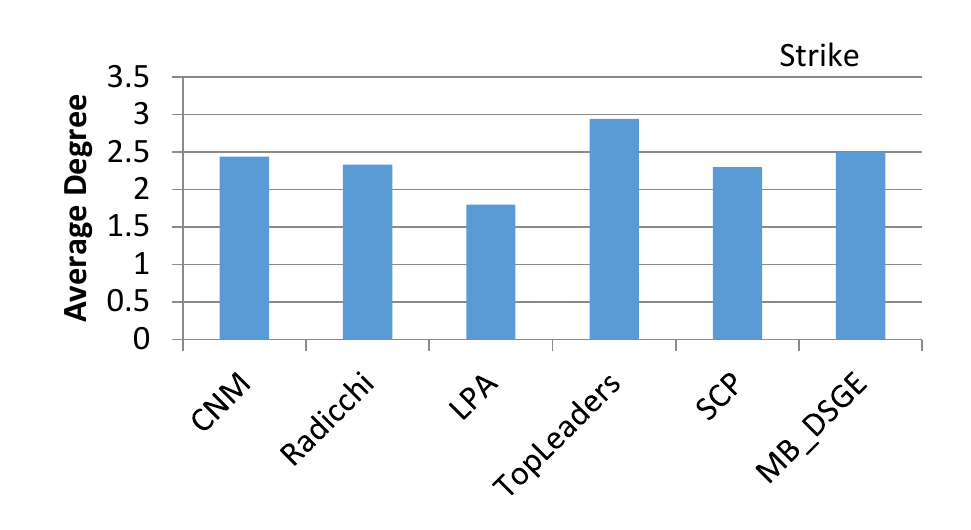}}%
	\subfigure[][]{%
	\centering
	\label{fig:strike_hepph_statsb}%
	\includegraphics[width=0.40\textwidth]{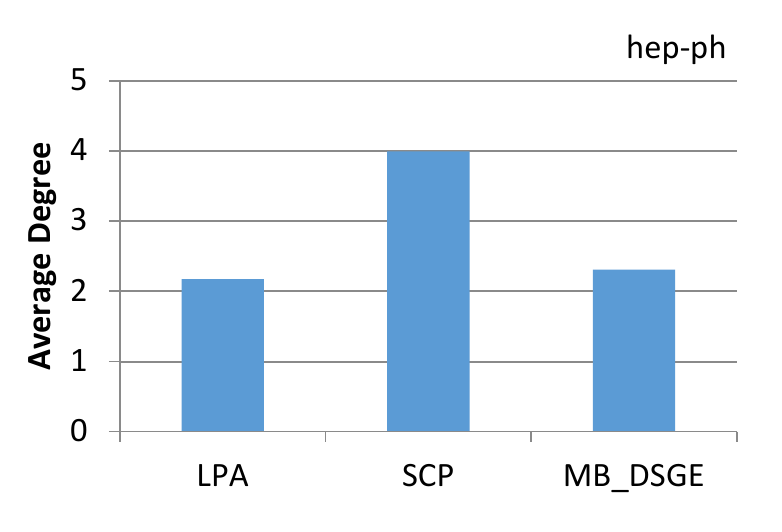}}%
	\\%
	\subfigure[][]{%
	\centering
	\label{fig:strike_hepph_statsc}%
	\includegraphics[width=0.5\textwidth]{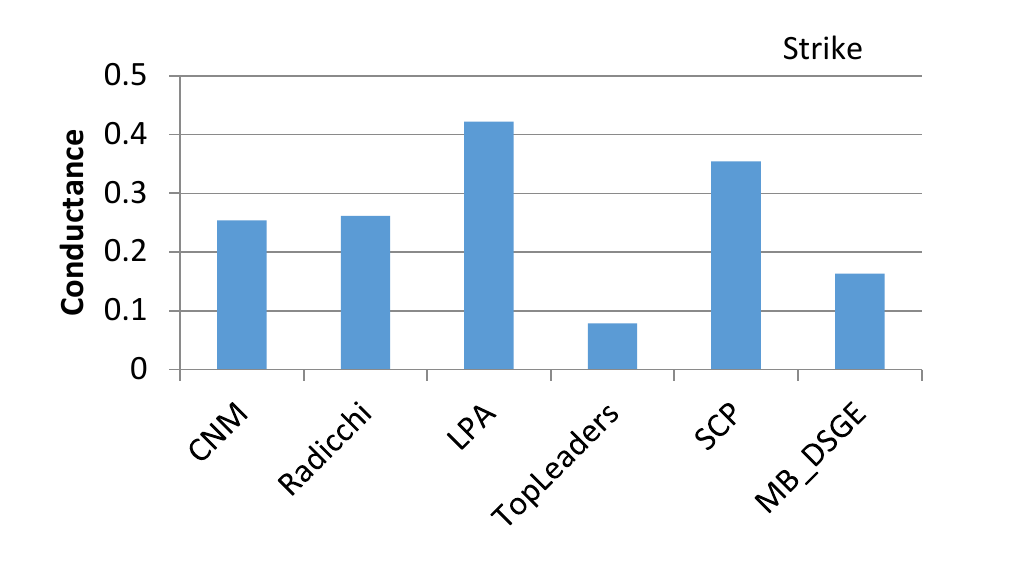}}%
	\subfigure[][]{%
	\centering
	\label{fig:strike_hepph_statsd}%
	\includegraphics[width=0.40\textwidth]{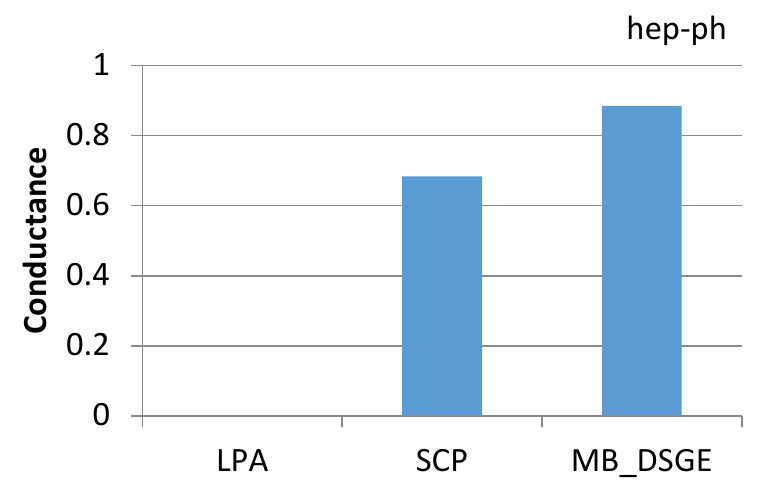}}%
	\\%
	\subfigure[][]{%
	\centering
	\label{fig:strike_hepph_statse}%
	\includegraphics[width=0.5\textwidth]{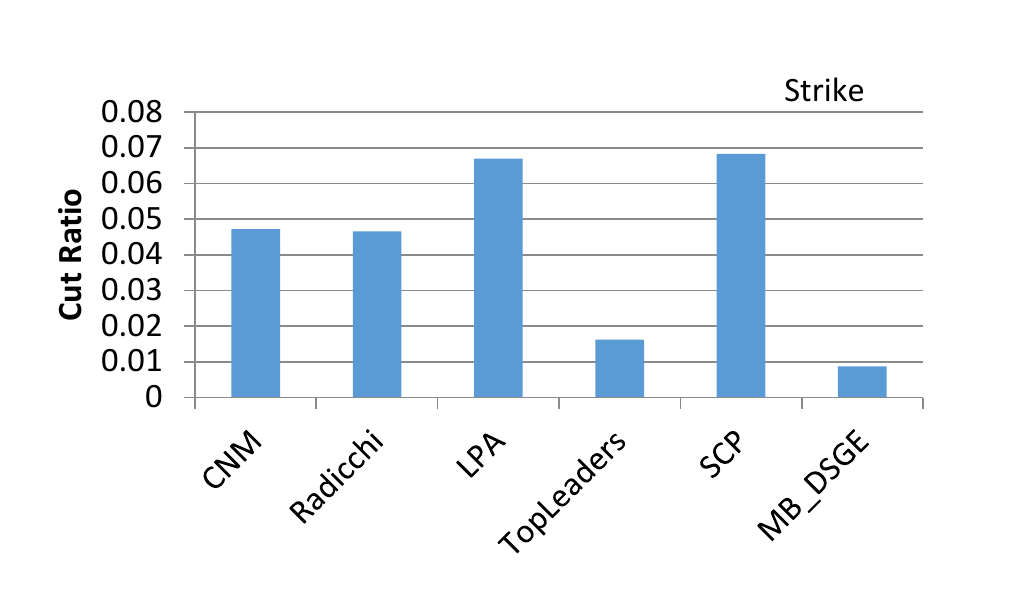}}%
	\subfigure[][]{%
	\centering
	\label{fig:strike_hepph_statsf}%
	\includegraphics[width=0.4\textwidth]{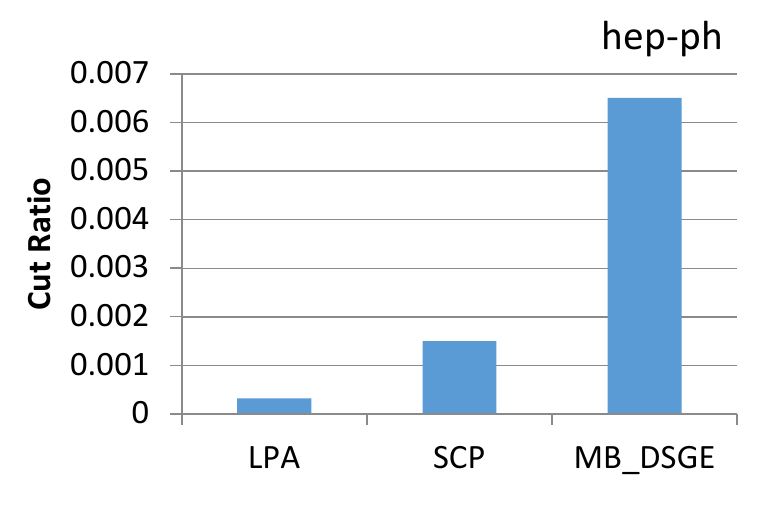}}%
	\caption{Analyzing the effectiveness of strike and hep-ph network communities respectively (a, b) average degree, (c, d) conductance, and (e, f) cut-ratio.}%
	\label{fig:strike_hepph_stats}%
\end{figure}

\subsection{Discussion}
The overall discussion in this section is carried out from two aspects: 1) analyzing different evaluation measures using same dataset, and 2) evaluating the clustering quality on datasets with varying properties.

\begin{figure}%
	\centering%
	\subfigure[][]{%
	\centering
	\label{fig:strike_statsa}%
	\includegraphics[width=0.5\textwidth]{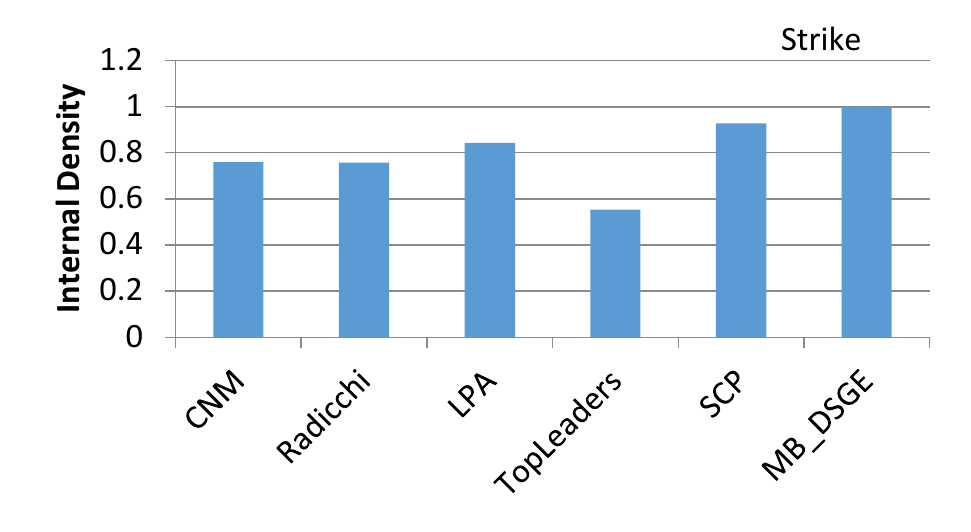}}%
	\subfigure[][]{%
	\centering
	\label{fig:strike_statsb}%
	\includegraphics[width=0.5\textwidth]{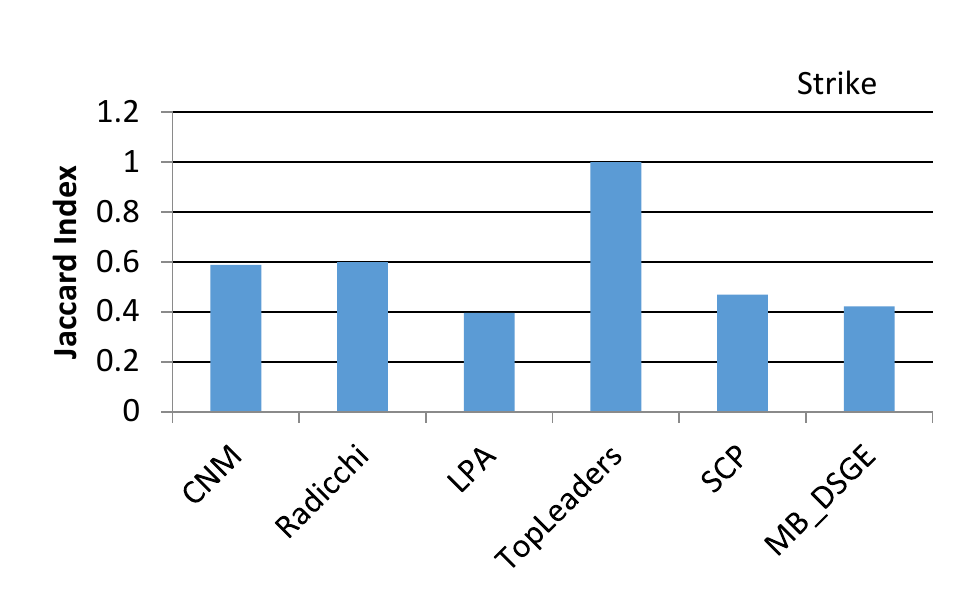}}%
	\\%
	\subfigure[][]{%
	\centering
	\label{fig:strike_statsc}%
	\includegraphics[width=0.5\textwidth]{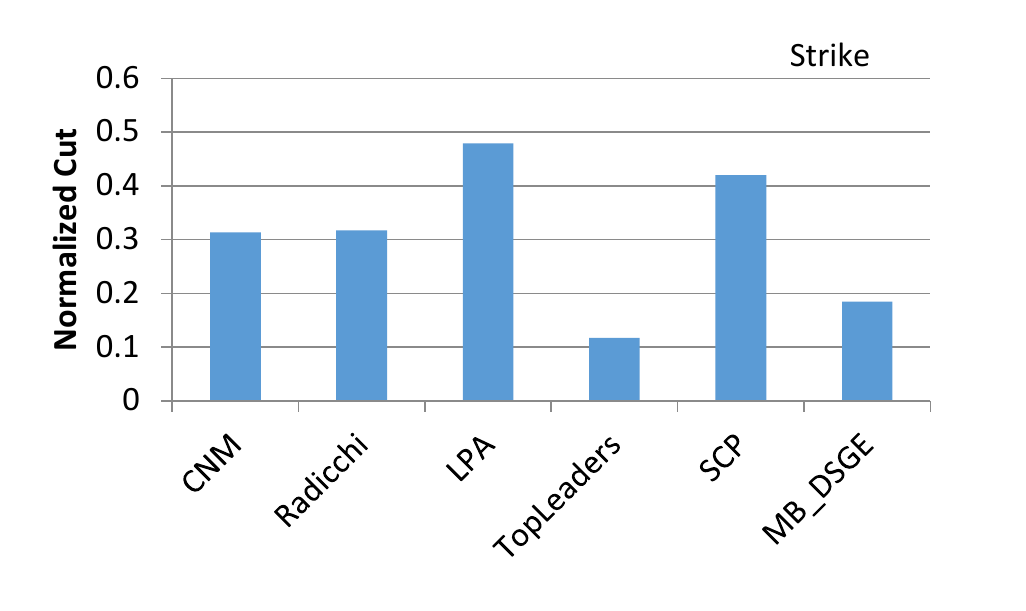}}%
	\subfigure[][]{%
	\centering
	\label{fig:strike_statsd}%
	\includegraphics[width=0.5\textwidth]{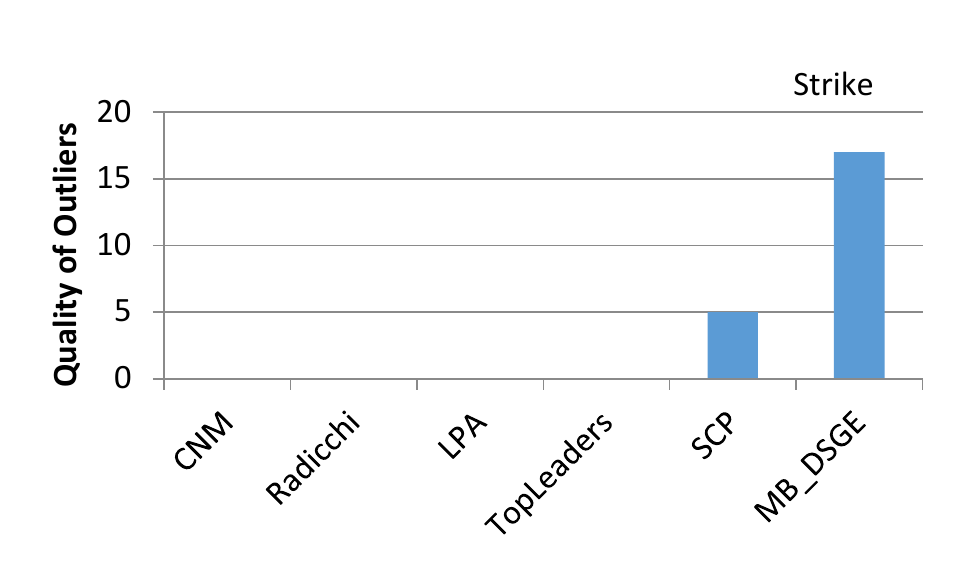}}%
    \caption{Analyzing the effectiveness, accuracy, and outlier scores for community detection approaches on strike network in terms of (a) internal density, (b) jaccard index, (c) normalized cut, and (d) outliers.}%
	\label{fig:strike_stats}%
\end{figure}

\begin{figure}
	\centering 
	\includegraphics[width=1.0\textwidth]{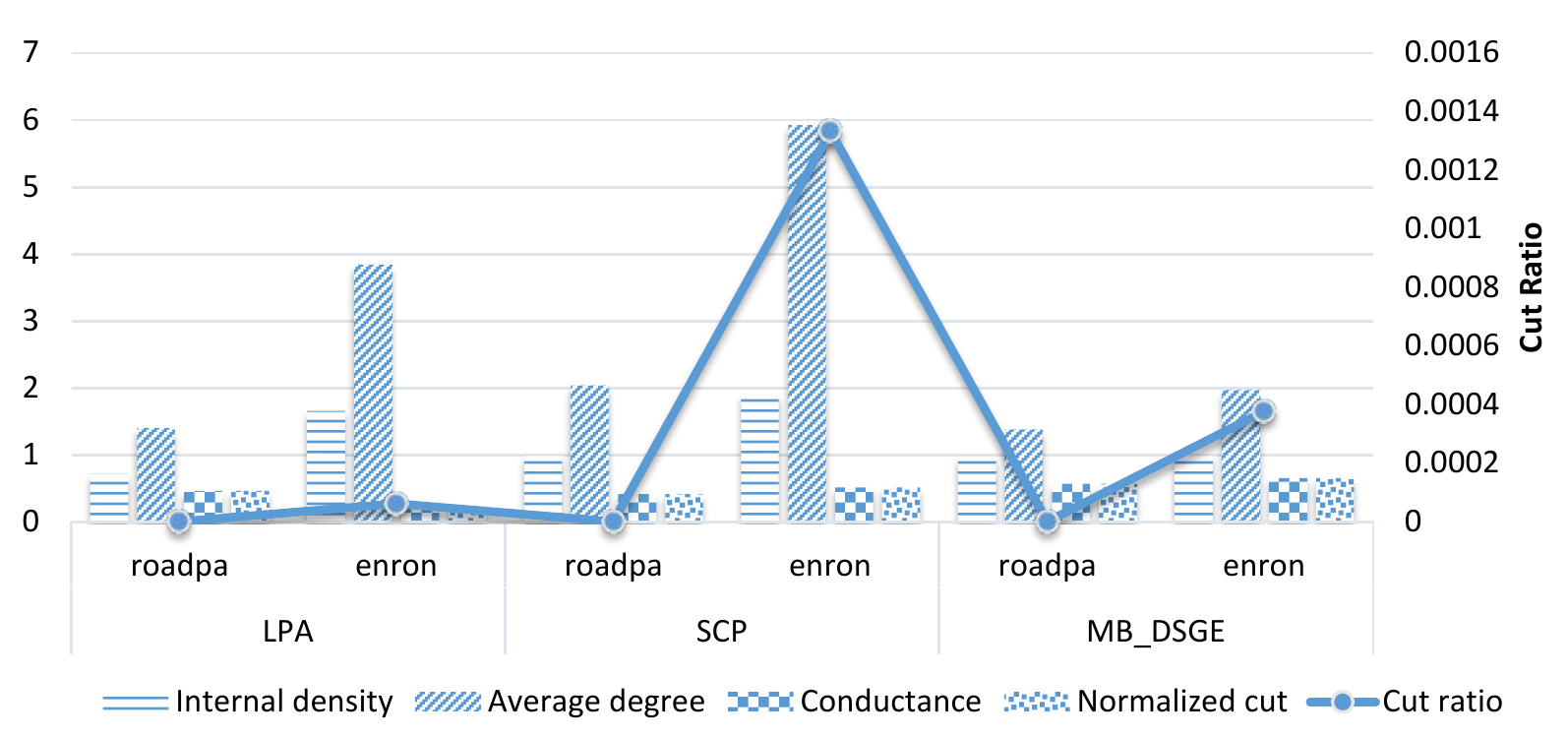}
	\caption{Analyzing the effectiveness of community detection approaches on Email and road-pa network in terms of internal density, average degree, conductance, normalized cut, and cut ratio.}
	\label{fig:measures_enron_roadpa}
\end{figure}

We analyzed the effectiveness of community detection methods on small-scale and medium-scale real life networks, i.e. Strike, Hep-ph, as shown in Figure \ref{fig:strike_hepph_stats} and Figure \ref{fig:strike_stats} respectively. The quality of communities is directly proportional to internal density and average degree values, i.e. good communities have higher values for these internal connectivity measures. In reference to existing study \cite{wang:becnhmark}, the LPA approach consistently outperforms other approaches through our experiments on Strike dataset, as depicted in Figure \ref{fig:strike_hepph_stats}. Similarly, MB-DSGE approach did not produce good quality clusters as claimed in the reference paper \textcolor{black}{\cite{wang:becnhmark}}. The only exception we observed for LPA is on Hep-ph dataset for average degree, where it has produced poor quality results compared with other methods\textcolor{black}{, as shown in Figure \ref{fig:strike_hepph_statsa} and Figure \ref{fig:strike_hepph_statsb}}.\textcolor{black}{In our understanding, as the connectivity among neighboring nodes is less in Hep-ph dataset compared with strike dataset, therefore, the same labels are not propagated to majority of the nodes. In other words, very few nodes collected same labels to be grouped together.}

\textcolor{black}{It is also important to note, in Figure \ref{fig:strike_hepph_stats}, that very few evaluation measures (community scoring function) may mislead our analysis. For instance, clique percolation approach shows community structures with increased average degree of the nodes but conductance and cut-ratio reveal different story. The reason is that for small size cliques it may result in good quality communities internally but when we analyze how good the community structure is externally then may be it is not. In other words, actual communities may not form clique structures, which is evident in Figure \ref{fig:strike} with actual communities as ground truth, and therefore this kind of conclusion can not be drawn from such inadequate analysis. We can conclude that average degree scoring function may not be a good choice in comparison to conductance and cut-ratio.}

\textcolor{black}{The strike and hep-ph datasets differ in size as well as in properties but not that different as road-network, however, we observe strange behavior for LPA, SCP and MS-DSGE algorithms. In Figure \ref{fig:strike_hepph_statse}, the communities generated through LPA and SCP have high cut-ratio compared with MS-DSGE. But in Figure \ref{fig:strike_hepph_statsf}, the effect is rather opposite for LPA and SCP. The community structure is already presented for Strike dataset and we can understand that due to better underlying community structures in the data, both LPA and SCP did not performed well. In other words, external links to other communities are too many that is why the resultant communities score is high in terms of cut-ratio. This is not the case with MS-DSGE as it find the dense subgraphs out of sparse graph. Is it the case with hep-ph dataset? In order to understand that we need to carefully look at the Figure \ref{fig:graph-nConnecta} where we plotted the number of neighbors and neighborhood connectivity. It clearly tells us that number of neighbors and the average neighborhood connectivity are complementary to each other. Therefore, LPA and SCP could perfom well, i.e. produced communities with low cut-ratio.}    

The higher conductance, normalized cut (internal and external connectivity), and cut ratio (external connectivity), the worse detected communities are. In other words, these measures are indirectly proportional to the quality of results. LPA shows good results, SCP is medium and MB-DSGE is bad on larger dataset, i.e. Hep-ph. The MB-DSGE algorithm is a method for identifying a set of dense subgraphs of a given sparse graph \cite{chen:MBDSGE}. Therefore, it has shown different results for these two datasets in Figure \ref{fig:strike_hepph_stats}, i.e. good results for Strike and relatively poor on Hep-ph dataset. 

The accuracy of communities is presented in Figure \ref{fig:strike_statsb} where ground-truth is available in advance. In this case, LPA algorithm has lower value for average Jaccard Index that is not same as in the base paper. However, other algorithms have shown the similar performance. Most likely these results deduced from the fact that the Strike dataset is relatively sparse and does not have outliers. 

The results quality in terms of outliers is also evaluated and shown in the Figure \ref{fig:strike_statsd}. The behavior of each algorithm under consideration is consistent with the claims in literature. MB-DSGE approach has produced the most outliers due to its natural instinct towards identifying the set of dense subgraphs on a given sparse graph. The outlier score for SCP method is relatively lower than MB-DSGE. 

We observed the effectiveness of LPA, SCP, and MB-DSGE algorithms on social and road networks, i.e. Enron and roadpa. Our objective is to investigate the variation in results towards diversified properties of these networks. We plotted the internal and external connectivity measures in Figure \ref{fig:measures_enron_roadpa}. The internal density and average degree for email (Enron) communities is higher than road (roadpa) communities and it is consistent for all three community detection methods. Since, road network is much sparse then email, therefore, they produced sparse communities with low average degree and density values. For other measures, including conductance, normalized cut, and cut ratio, LPA has outperformed SCP and MB-DSGE methods. The SCP method achieved higher score for internal density and average degree on both datasets compared with LPA, which is non-trivial. However, the accuracy aspect is compromised for SCP. There is a trade-off between effectiveness and accuracy when it comes to SCP method.

\textcolor{black}{It is interesting to see that LPA, SCP and MB-DSGE methods detected better communities in terms of cut-ratio on road network in Figure \ref{fig:measures_enron_roadpa}, where the road network is comparatively sparse than Enron email network. It make sense because in road network the detected communities have less number of edges going out of the communities due to sparseness, which is not the case with Enron email network. For other measures like conductance and normalize-cut, we can see the similar behavior of the algorithms on read network where communities have relatively lower values compared with the Enron network. There is an exception for LPA algorithm when it comes to road network where it has produced relatively poor communities in reference to email network. The reason for such behavior lies in label propagation algorithm where strongly connected nodes end up receiving same labels and resulted in the same community. In other words, it does not include the nodes in the same community having links to nodes of other communities.}

\section{Conclusion and Future Directions}
This study provided an experimental evaluation of a set of representative and well-known community detection algorithms on structurally different datasets with varying properties, i.e. density, sparsity, and neighborhood connectivity. We evaluated results of these algorithms for effectiveness, accuracy, and outliers. The extensive evaluation of the resultant communities in terms of unique measures suggested the superiority of LPA method over others in social networks. However, SCP method achieved better internal density and average degree compared to LPA, while with a slight compromise on accuracy. The impact of network properties is proportionally reflected in results, but the behavior of community detection methods persisted.

It is non-trivial to foresee the impact of community detection approaches over weighed and directed graphs. The personal traits of entities in a network, e.g. attributes associated with the vertices, further complicates this process and requires a systematic analysis for better understanding. \textcolor{black}{The analysis problem becomes even more complex when we encounter non-homogeneous networks with varying types of vertices, i.e. heterogeneous networks.}

\section*{Acknowledgement}
I acknowledge the support of Islamic University of Madinah for disseminating this research work. I also appreciate the help provided by Zilya Yagafarova and support of colleagues at Innopolis University Russia to carryout this work.

\balance
\bibliographystyle{IEEEtran}
\bibliography{references}  

\begin{thebibliography}{10}
\providecommand{\url}[1]{#1}
\csname url@samestyle\endcsname
\providecommand{\newblock}{\relax}
\providecommand{\bibinfo}[2]{#2}
\providecommand{\BIBentrySTDinterwordspacing}{\spaceskip=0pt\relax}
\providecommand{\BIBentryALTinterwordstretchfactor}{4}
\providecommand{\BIBentryALTinterwordspacing}{\spaceskip=\fontdimen2\font plus
\BIBentryALTinterwordstretchfactor\fontdimen3\font minus
  \fontdimen4\font\relax}
\providecommand{\BIBforeignlanguage}[2]{{%
\expandafter\ifx\csname l@#1\endcsname\relax
\typeout{** WARNING: IEEEtran.bst: No hyphenation pattern has been}%
\typeout{** loaded for the language `#1'. Using the pattern for}%
\typeout{** the default language instead.}%
\else
\language=\csname l@#1\endcsname
\fi
#2}}
\providecommand{\BIBdecl}{\relax}
\BIBdecl

\bibitem{clauset:CNM}
A.~Clauset, M.~E.~J. Newman, and C.~Moore, ``Finding community structure in
  very large networks,'' \emph{Phys. Rev. E}, vol.~70, no.~6, p. 066111, 2004.

\bibitem{chen:MBDSGE}
J.~Chen and Y.~Saad, ``Dense subgraph extraction with application to community
  detection,'' \emph{Knowledge and Data Engineering, IEEE Transactions on},
  vol.~24, no.~7, pp. 1216--1230, 2012.

\bibitem{huang:gcluskeleton}
J.~Huang, H.~Sun, Q.~Song, H.~Deng, and J.~Han, ``Revealing density-based
  clustering structure from the core-connected tree of a network,''
  \emph{Knowledge and Data Engineering, IEEE Transactions on}, vol.~25, no.~8,
  pp. 1876--1889, 2013.

\bibitem{khorasgani:topleaders}
R.~R. Khorasgani, J.~Chen, and O.~R. Za{\"\i}ane, ``Top leaders community
  detection approach in information networks,'' in \emph{4th SNA-KDD Workshop
  on Social Network Mining and Analysis}.\hskip 1em plus 0.5em minus
  0.4em\relax Citeseer, 2010.

\bibitem{kumpula:SCP}
J.~M. Kumpula, M.~Kivel{\"a}, K.~Kaski, and J.~Saram{\"a}ki, ``Sequential
  algorithm for fast clique percolation,'' \emph{Physical Review E}, vol.~78,
  no.~2, p. 026109, 2008.

\bibitem{leung:HANP}
I.~X. Leung, P.~Hui, P.~Lio, and J.~Crowcroft, ``Towards real-time community
  detection in large networks,'' \emph{Physical Review E}, vol.~79, no.~6, p.
  066107, 2009.

\bibitem{raghavan:LPA}
U.~N. Raghavan, R.~Albert, and S.~Kumara, ``Near linear time algorithm to
  detect community structures in large-scale networks,'' \emph{Physical Review
  E}, vol.~76, no.~3, p. 036106, 2007.

\bibitem{zhao:MKMF}
F.~Zhao and A.~K. Tung, ``Large scale cohesive subgraphs discovery for social
  network visual analysis,'' \emph{Proceedings of the VLDB Endowment}, vol.~6,
  no.~2, pp. 85--96, 2012.

\bibitem{yang:defining}
J.~Yang and J.~Leskovec, ``Defining and evaluating network communities based on
  ground-truth,'' \emph{Knowledge and Information Systems}, vol.~42, no.~1, pp.
  181--213, 2015.

\bibitem{xie2013overlapping}
J.~Xie, S.~Kelley, and B.~K. Szymanski, ``Overlapping community detection in
  networks: The state-of-the-art and comparative study,'' \emph{Acm computing
  surveys (csur)}, vol.~45, no.~4, p.~43, 2013.

\bibitem{harenberg:lpaevaluation}
S.~Harenberg, G.~Bello, L.~Gjeltema, S.~Ranshous, J.~Harlalka, R.~Seay,
  K.~Padmanabhan, and N.~Samatova, ``Community detection in large-scale
  networks: a survey and empirical evaluation,'' \emph{Wiley Interdisciplinary
  Reviews: Computational Statistics}, vol.~6, no.~6, pp. 426--439, 2014.

\bibitem{leskovec:empirical}
J.~Leskovec, K.~J. Lang, and M.~Mahoney, ``Empirical comparison of algorithms
  for network community detection,'' in \emph{Proceedings of the 19th
  international conference on World wide web}.\hskip 1em plus 0.5em minus
  0.4em\relax ACM, 2010, pp. 631--640.

\bibitem{wang:becnhmark}
M.~Wang, C.~Wang, J.~X. Yu, and J.~Zhang, ``Community detection in social
  networks: An in-depth benchmarking study with a procedure-oriented
  framework,'' \emph{PVLDB}, vol.~8, no.~10, pp. 998--1009, 2015.

\bibitem{misra2016non}
G.~Misra, J.~M. Such, and H.~Balogun, ``Non-sharing communities? an empirical
  study of community detection for access control decisions,'' in \emph{2016
  IEEE/ACM International Conference on Advances in Social Networks Analysis and
  Mining (ASONAM)}.\hskip 1em plus 0.5em minus 0.4em\relax IEEE, 2016, pp.
  49--56.

\bibitem{garg2017comparative}
N.~Garg and R.~Rani, ``A comparative study of community detection algorithms
  using graphs and r,'' in \emph{2017 International Conference on Computing,
  Communication and Automation (ICCCA)}.\hskip 1em plus 0.5em minus 0.4em\relax
  IEEE, 2017, pp. 273--278.

\bibitem{zhao2018comparative}
Z.~Zhao, S.~Zheng, C.~Li, J.~Sun, L.~Chang, and F.~Chiclana, ``A comparative
  study on community detection methods in complex networks,'' \emph{Journal of
  Intelligent \& Fuzzy Systems}, no. Preprint, pp. 1--10, 2018.

\bibitem{EmpStdCommunityDetect}
K.~Chandusha, S.~R. Chintalapudi, and M.~H.~M. Krishna~Prasad, ``An empirical
  study on community detection algorithms,'' in \emph{Smart Intelligent
  Computing and Applications}, S.~C. Satapathy, V.~Bhateja, and S.~Das,
  Eds.\hskip 1em plus 0.5em minus 0.4em\relax Singapore: Springer Singapore,
  2019, pp. 35--44.

\bibitem{radicchi:defining}
F.~Radicchi, C.~Castellano, F.~Cecconi, V.~Loreto, and D.~Parisi, ``Defining
  and identifying communities in networks,'' \emph{Proceedings of the National
  Academy of Sciences of the United States of America}, vol. 101, no.~9, pp.
  2658--2663, 2004.

\bibitem{adamcsek2006cfinder}
B.~Adamcsek, G.~Palla, I.~J. Farkas, I.~Der{\'e}nyi, and T.~Vicsek, ``Cfinder:
  locating cliques and overlapping modules in biological networks,''
  \emph{Bioinformatics}, vol.~22, no.~8, pp. 1021--1023, 2006.

\bibitem{urlcode:urlcnm}
A.~Clauset, ``Source code of cnm algorithm,''
  \url{http://www.cs.unm.edu/~aaron/research/fastmodularity.htm}
  \textcolor{black}{(Last accessed on 16 March, 2019)}.

\bibitem{urlcode:urlradicchi}
F.~Radicchi, ``Source code of radicchi algorithm,''
  \url{http://homes.soic.indiana.edu/filiradi/resources.html}
  \textcolor{black}{(Last accessed on 16 March, 2019)}.

\bibitem{urlcode:urllpar}
``R code of lpa algorithm,''
  \url{http://igraph.wikidot.com/community-detection-in-r}
  \textcolor{black}{(Last accessed on 16 March, 2019)}.

\bibitem{urlcode:urllpapython}
``Python code of lpa algorithm,''
  \url{http://orkohunter-networkx.readthedocs.org/en/latest/_modules/networkx/algorithms/community/asyn_lpa.html}
  \textcolor{black}{(Last accessed on 16 March, 2019)}.

\bibitem{urlcode:urltopleaders}
R.~R. Khorasgani, ``Source code of topleaders algorithm,''
  \url{https://webdocs.cs.ualberta.ca/~rabbanyk/TopLeader/}
  \textcolor{black}{(Last accessed on 16 March, 2019)}.

\bibitem{urlcode:urlscp}
J.~M. Kumpula, ``Source codes of scp algorithm,''
  \url{http://complex.cs.aalto.fi/resources/software/} \textcolor{black}{(Last
  accessed on 16 March, 2019)}.

\bibitem{urlcode:urlmbdsge}
``Code of mbdsge algorithm,''
  \url{https://github.com/sranshous/Graph-Community-Detection}
  \textcolor{black}{(Last accessed on 16 March, 2019)}.

\bibitem{urlcode:eigen}
``Eigen: Linear algebra library,''
  \url{http://eigen.tuxfamily.org/index.php?title=Main_Page}
  \textcolor{black}{(Last accessed on 16 March, 2019)}.

\bibitem{snapnets}
J.~Leskovec and A.~Krevl, ``{SNAP Datasets}: {Stanford} large network dataset
  collection,'' \url{http://snap.stanford.edu/data} \textcolor{black}{(Last
  accessed on 16 March, 2019)}, 2014.

\bibitem{leskovec2008statistical}
J.~Leskovec and et. al., ``Statistical properties of community structure in
  large social and information networks,'' in \emph{Proceedings of the 17th
  international conference on World Wide Web}.\hskip 1em plus 0.5em minus
  0.4em\relax ACM, 2008, pp. 695--704.

\bibitem{leskovec2009community}
J.~Leskovec, K.~J. Lang, A.~Dasgupta, and M.~W. Mahoney, ``Community structure
  in large networks: Natural cluster sizes and the absence of large
  well-defined clusters,'' \emph{Internet Mathematics}, vol.~6, no.~1, pp.
  29--123, 2009.

\bibitem{shi2000normalized}
J.~Shi and J.~Malik, ``Normalized cuts and image segmentation,'' \emph{Pattern
  Analysis and Machine Intelligence, IEEE Transactions on}, vol.~22, no.~8, pp.
  888--905, 2000.

\bibitem{fortunato2010community}
S.~Fortunato, ``Community detection in graphs,'' \emph{Physics Reports}, vol.
  486, no.~3, pp. 75--174, 2010.

\end{thebibliography}

\end{document}